# Temporal resolution enhancement in Structured Illumination Microscopy using cascaded reconstruction

Doron Shterman,[1] Guy Bartal[1,*]

[1]*The Erna and Andrew Viterbi Faculty of Electrical and computer Engineering, Technion – Israel Institute of Technology, Haifa 32000, Israel*
*[guy@ee.technion.ac.il](guy@ee.technion.ac.il)*

**Structured Illumination Microscopy (SIM) allows access to spatial information beyond the diffraction limit by folding high frequency components into the optical system's baseband. Using various algorithmic techniques, an image containing sub-wavelength information can be reconstructed. While linear SIM is considered superior to other Super-Resolution (SR) methods in its compatibility with live cell imaging and optical setup simplicity, it is inherently limited in terms of its temporal resolution as each image requires multiple frames. Here we present a practical and efficient reconstruction approach supporting up to 3-fold temporal resolution increase with SIM, using overlapping regions in the folded frequency components within the Fourier domain. Our approach can be readily implemented in any previously introduced SIM realization to both improve the temporal resolution and to simplify the optical apparatus.**

The ultimate temporal resolution limit in super-resolution (SR) fluorescence microscopy is the fluorophore excitation-emission cycle. However, this limit has never been reached in practice owing to system-level constraints. Among these constraints, we can include the physical apparatus, concept of operation, detector efficiency, and image acquisition time. In single-molecule localization methods, such as STORM [1] and PALM [2], the temporal resolution is set by the need to record sufficient number of images to accurately locate single molecules and reveal their biological structures. In patterned-illumination methods, such as parallelized STED [3] and SIM [4], the temporal resolution is limited by the scanning duration (OL-STED [5]) or by the need to acquire multiple frames with different illumination profiles in order to resolve frequency aliasing, per each reconstructed super-resolved image. Various strategies have been employed to improve the temporal resolution in SR microscopy [6,7]; however, the trade-off between spatial and temporal resolution remains.

Currently, SIM is considered to be among the fastest SR microscopy techniques [8,9] because of its superior photon efficiency, allowing most of the photons collected to be used for super-resolved image formation. Reducing the overall number of frames captured for the reconstruction can improve the temporal resolution in SIM as well as reduce phototoxicity or bleaching effects, to support live cell SR imaging. This becomes a dire need for SIM realizations that support spatial resolution enhancement beyond the classical 2-fold improvement. In MM-SIM [9,10], for example, an up to 5-fold increase in spatial resolution can be achieved, yet with a staggering 72 frames per single super-resolved image. Methods introduced in the past to reduce data capture in SIM rely however on complicated reconstruction schemes [12–14] with high sensitivity to the reconstruction parameters, resulting in reconstruction errors and artifacts.

Here, we propose an efficient and practical super-resolution image reconstruction approach capable of improving the temporal resolution of SIM up to 3-fold. The reconstruction efficiency is based on frequency de-aliasing with a reduced number of frames. This is achieved by exploiting the overlapping regions in the Fourier domain, where both the known and unknown spectral components are mixed together. In contrast to previously suggested reconstruction methods [13], the proposed increase in temporal resolution is achieved with zero dependency between frames captured in different illumination orientations. In addition, reconstruction can be performed without relying on the complicated curvilinear polygonal regions in the Fourier domain.

The optical resolution limit is conveyed in the Fourier domain as the cutoff frequency $k_0$ (Fig. 1.a), which is the maximum spectral frequency that can be resolved using an optical imaging system.

$$k_0 = \frac{2NA}{\lambda} = \frac{2n \cdot sin(\alpha)}{\lambda} \qquad (1)$$

where $\lambda$ is the wavelength of light, NA is the numerical aperture of the imaging system, $n$ is the index of refraction, and $\alpha$ is the half-angle of the maximum cone of light that can enter or exit the lens.

Conceptually, the diffraction limit can be described as a low-pass filter in the Fourier domain, through which only low frequencies below the cutoff frequency can pass. Any information encompassed in frequencies beyond this limit cannot be resolved, and an image containing such high spectral frequencies eventually appears blurred or has indistinguishable features. Thus, in practice, patterned-illumination methods, such as SIM, expand the Fourier domain content of a diffraction-limited image to reconstruct a super-resolved image.

Recalling the SIM mathematical framework, an illumination profile generated with two coherent beams takes the following form:

$$I_{ill}(r) = I_0\left[1 + \frac{\cos}{2}(2\pi p \cdot r + \varphi)\right] \quad (2)$$

where $r \equiv (x, y)$ is the two-dimensional spatial position vector, $I_0$ is the peak illumination intensity taken as unity for simplicity from this point forward, $p$ is the illumination frequency vector in the Fourier domain, and $\varphi$ is the phase of the illumination profile.

In Fourier domain, the Illumination signal is given by:

$$\tilde{I}_{ill}(k) = \delta(k) + e^{i\varphi}\frac{\delta}{2}(k+p) + e^{-i\varphi}\frac{\delta}{2}(k-p) \quad (3)$$

where ~ denotes the Fourier transform, $\delta$ denotes the Dirac delta function, and $k \equiv (k_x, k_y)$ denotes the two-dimensional Fourier domain position vector.

According to Fourier theorem, the emitted light is given by:

$$\tilde{I}_{em}(k) = \tilde{s}(k) \otimes \tilde{I}_{ill}(k) \quad (4)$$

where $\tilde{s}(k)$ is the fluorophore density distribution in the Fourier domain and $\otimes$ denotes the convolution operator.

The detected signal is given by:

$$\tilde{I}_{dec}(k) = \tilde{I}_{em}(k) \cdot \tilde{h}(k) = \left[\tilde{s}(k) \otimes \tilde{I}_{ill}(k)\right] \cdot \tilde{h}(k) \quad (5)$$

where · denotes point-wise multiplication and $\tilde{h}(k)$ is the system's Optical Transfer Function (OTF).

Combining Eq. (3) and Eq. (5) for a given illumination profile phase $\varphi_m$, the detected signal in the Fourier domain is given by:

$$\tilde{I}_{dec}(k) = \left[\tilde{s}(k) + e^{i\varphi_m}\frac{\tilde{s}}{2}(k+p) + e^{-i\varphi_m}\frac{\tilde{s}}{2}(k-p)\right] \cdot \tilde{h}(k) \quad (6)$$

Equation (6) shows that the detected signal $\tilde{I}_{dec}(k)$ contains two distinct shifted and one non-shifted fluorophore spectral component folded into the OTF passband region. By capturing a set of three consequent images while changing the illumination profile phase $\varphi_m$ in between, one can de-alias the shifted spectral components and reconstruct $\tilde{s}$ for all frequencies shifted into, or originally included within, the OTF passband. This can be accomplished by solving a set of three linear equations:

$$\begin{bmatrix}\tilde{I}_{dec,1}(k)\\\tilde{I}_{dec,2}(k)\\\tilde{I}_{dec,3}(k)\end{bmatrix} = \left[\begin{pmatrix}2 & e^{i\varphi_1} & e^{-i\varphi_1}\\2 & e^{i\varphi_2} & e^{-i\varphi_2}\\2 & e^{i\varphi_3} & e^{-i\varphi_3}\end{pmatrix}\begin{bmatrix}\tilde{s}(k)\\\tilde{s}(k+p)\\\tilde{s}(k-p)\end{bmatrix}\right] \cdot \frac{\tilde{h}(k)}{4} \quad (7)$$

With the inversion of the 3X3 phase matrix, as shown in Eq. (7), the three spectral components $\tilde{s}(k)$, $\tilde{s}(k+p)$ and $\tilde{s}(k-p)$ are retrieved. Next, the extended Fourier representation is realized by placing each spectral component back into its original position while removing redundant overlaps. This process is then repeated for each illumination profile frequency $p$, to complete the continuous reconstruction of the Fourier domain. Thus, de-aliasing, or spectral component unmixing, is the first processing stage in most SIM reconstruction schemes.

Alternatively, we suggest exploiting the spatial distribution of frequency content in the Fourier domain to perform the required de-aliasing. For ease of annotation, we define $\tilde{s}_-$, $\tilde{s}_+$ and $\tilde{s}_0$ as the shifted spectral components $\tilde{s}(k-p)$, $\tilde{s}(k+p)$ and the unshifted component $\tilde{s}(k)$ accordingly. In addition, we use the annotations L and R to denote the spectral component content on the Left- or Right-hand side of the Fourier domain around a translation axis of choice $v$, originating at the origin of the Fourier domain. A geometrical illustration of the shifted and unshifted spectral components, after or before aliasing, is shown in Fig. 1. In this example we choose $v \equiv y$ and $\tilde{s}_{ij}$ are defined with $i = [0, -, +]$, and $j = [L, R]$.

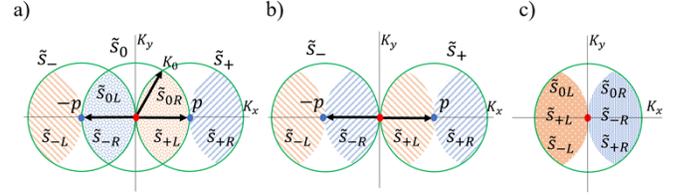

Fig. 1. **Geometrical illustration of SIM aliasing in Fourier domain**. $k_0$ is the diffraction cutoff frequency, $p$ is the illumination profile frequency, chosen to be equal in size to $k_0$. (a) 1D spectral components in SIM, prior to aliasing. The three distinct spectral regions, being the DC and two shifted components, are marked with green circle shape with cutoff frequency of $k_0$ corresponding to the 2D OTF shape. The two shifted components are shifted along the x direction with the maximal shift possible $|p| = |k_0|$ when using the same optical system for both illumination and transmission paths. Blue and orange strips are contained within an ellipse-like shape corresponding to the overlapping regions between the DC and shifted spectral components. (b) Shifted spectral components without the overlapping DC component, showing the folded spectral regions in their original position in Fourier domain before aliasing. (c) Spectral components position post aliasing. Ellipse-like shapes with blue and orange colors contain shifted and unshifted spectral content within the overlapping region.

Once captured through an optical system, aliasing effectively mixes shifted and unshifted spectral components. The overlapping components are contained within the intersection region of the circular OTF shapes, that is, an ellipse-like region. Six distinct spectral components are contained within these regions, as shown in Fig. 1.c. Three components: $\tilde{s}_{-L}, \tilde{s}_{+L}, \tilde{s}_{0L}$ residing on the left-hand side of the y-axis, and another three: $\tilde{s}_{-R}, \tilde{s}_{+R}, \tilde{s}_{0R}$ residing on the right-hand side of the y-axis.

Since the Fourier transform of a real-valued signal obeys conjugation symmetry, the Fourier transform of $I_{dec}(r)$, denoted as $\tilde{I}_{dec}$, follows the equation:

$$\tilde{I}_{dec}(k) = \tilde{I}_{dec^*}(-k) \quad (8)$$

where * denotes complex conjugate operator.

Hence, instead of decoding all six spectral components $\tilde{s}_{ij}$, conjugation symmetry supports spectral unmixing by decoding only three spectral components. In classical linear SIM, this is achieved with three consequent image captures per folded spectral content, generating three independent linear equations.

Taking a closer look at the geometrical representation of aliasing in Fig. 1, before aliasing occurs, we notice additional relationships between the depicted spectral components:

$$\tilde{s}_{0L} = \tilde{s}_{-R} \qquad (9)$$

$$\tilde{s}_{0R} = \tilde{s}_{+L} \qquad (10)$$

By capturing a single diffraction-limited image under uniform illumination, we can directly measure $\tilde{s}_{0L}$ and $\tilde{s}_{0R}$. Hence, out of the three components to be decoded, such as those to the right-hand side of the y-axis: $\tilde{s}_{-R}, \tilde{s}_{+R}, \tilde{s}_{0R}$, by capturing just two images instead of three, there is enough information to complete the entire decoding process. Defining the right-hand side of an aliased image in the Fourier domain as $\tilde{I}_{dec,R}(k) \equiv \tilde{s}_{-R} + \tilde{s}_{+R} + \tilde{s}_{0R}$, and using the DC components $\tilde{s}_{0L}$ and $\tilde{s}_{0R}$, we can perform the following substraction denoted by $\tilde{I}_{sub,R}(k)$:

$$\tilde{I}_{sub,R}(k) = \tilde{I}_{dec,R}(k) - \tilde{s}_{0R} - \tilde{s}_{0L} \qquad (11)$$

Substituting Eq. (11) with Eq. (10) and Eq. (9) we get:

$$\tilde{I}_{sub,R}(k) = \tilde{s}_{+R} \qquad (12)$$

Finally, using two image captures instead of three, and without any prior knowledge of the phase of the illumination profile, we can recover the mixed spectral components and reconstruct the Fourier domain representation of a super-resolved image. This de-aliasing scheme is then repeated for each folded frequency using an overlap with known spectral content, originating from either the DC image or from previously decoded spectral content. Theoretically, this can be expanded in a cascading manner to support unlimited [11,15] spatial resolution enhancement, where the spectral components are resolved sequentially from low to high frequencies. In the cascading reconstruction flow, the known DC components used in Eq. (11), are replaced with the recovered spectral components to support the unmixing of the higher-frequency content. For n folded frequencies, a total of n+1 images are required, compared to the 3·n required in linear SIM, representing up to ~3 times improvement in data collection efficiency or, equivalently, a ~3 times increase in temporal resolution.

In practice, an ellipse-like shape limits the useful overlapping regions. When employed for linear SIM, where the illumination frequency vector $p$ is equal to $k_0$, some regions in the Fourier domain cannot be recovered, as shown in Fig. 2.d-f. In this case, the reconstructed super-resolved image is expected to contain distortions owing to the lack of continuous resolution enhancement in all spatial orientations. This can be resolved by either decreasing the size of the illumination frequency vector $p$ or by increasing the number of spatial orientations recovered. These resolutions are tightly coupled; the smaller $p$ vector, the lesser the number of spatial orientations needed to achieve continuous coverage (see Fig. 2.g-i). Alternatively, the missing Fourier components can also be recovered using previously proposed algorithmic methods [12,13], without the need to capture additional frames.

Compared to classical SIM reconstruction, the de-aliasing mechanism proposed herein allows significant temporal resolution enhancement with little to no effect on the spatial resolution. Compared with previously proposed reconstruction methods that rely on a reduced number of frames, the proposed mechanism offers a simplified implementation process in addition to completely relieving interdependencies between the captured moiré fringes. This means that the instantaneous temporal resolution, per each single spatial orientation, is improved by an additional 3-fold factor accounting for a total of 9-fold compared to the classical SIM reconstruction approach.

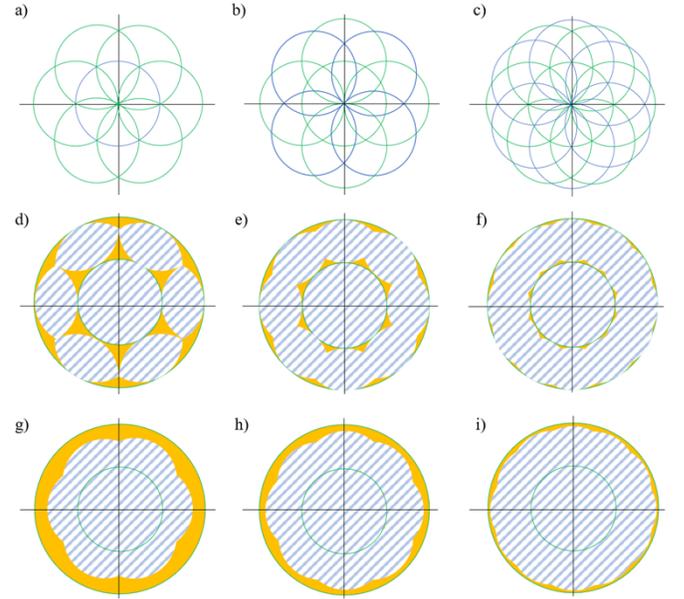

Fig. 2. **SIM reconstruction comparison**. (a-c) Reconstructed effective OTF in classical linear SIM with 3, 4 and 6 evenly distributed spatial orientations respectively. Circular regions in green and blue resemble the system's OTF baseband shape. (d-f) Equivalent reconstructed OTF, with the proposed reconstruction approach, achieved with 3-fold less frames. Both the DC components and the reconstructed spectral components are marked with blue strips. The orange circle in the background shows the maximum Fourier region that can be reconstructed with classical SIM. Hence, Fourier regions shown in orange indicates an unreconstructed portion of the Fourier domain. As the amount of recovered spatial orientations increases, from 3 to 6 in this example, the less unreconstructed regions remain. (g-i) Slight reduction in the maximal reconstructed spectral frequency, achieved by defining the illumination frequency $|p| < |k_0|$, allows continuous reconstruction of Fourier domain using the proposed reconstruction approach.

By exploiting the entire ellipse-like overlapping region in the Fourier domain, we can recover the highest span of aliased spectral components. However, because most modern image sampling devices and digital image representations use rectangular pixel arrays, employing curvilinear polygonal regions may lead to reconstruction errors owing to mismatched pixels along the curved boundaries. Alternatively, polygonal shapes can be cropped out from the ellipse-like region and significantly improve the reconstruction results, at the cost of slight reduction in the spatial resolution.

For simulation purposes, we used low-resolution, diffraction-limited, and high-resolution reconstructed SIM images of a biological sample, as shown in Fig. 3. Using the entire ellipse-like overlapping region, we demonstrate cascading reconstruction up to a maximal frequency of $|k_f| = |2k_i + k_0|$, where $|k_i| = \frac{8}{9}|k_0|$ and $k_0$ is the cutoff frequency dictated by an airy disk PSF.

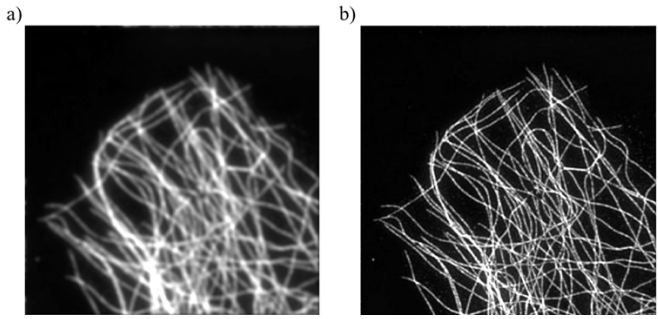

Fig. 3. **U2OS cells data used for simulation**. The cells were fixed in PHEMS\glutaraldehyde, labeled with DM1 alpha anti-Tubulin primary antibody and Alexa 488 secondary. (a) Wide-field U2OS cells image. (b) Super-resolved image generated with SIM. Sample data used with permission of the Department of Cell Biology at Harvard Medical School.

To demonstrate the cascading nature of the proposed reconstruction we generated 10 distinct illumination profiles, each containing a discrete frequency $k_i = \frac{8}{9}|k_0|\beta\theta_i$, where $\beta\theta_i \in [1 \cdot (-45°, 0°, 45°, 90°), 2 \cdot (-60°, -30°, 0°, 30°, 60°, 90°)]$ and a single uniform illumination profile, required to capture the unperturbed DC component. The ellipse-like Fourier regions, marked in Red in Fig. 4, were decoded as described in Eq. (11), relocated to their original position in the Fourier domain, and merged using a Wiener filter, as suggested in the classical SIM reconstruction [4], to create a continuous reconstruction of the image Fourier domain representation.

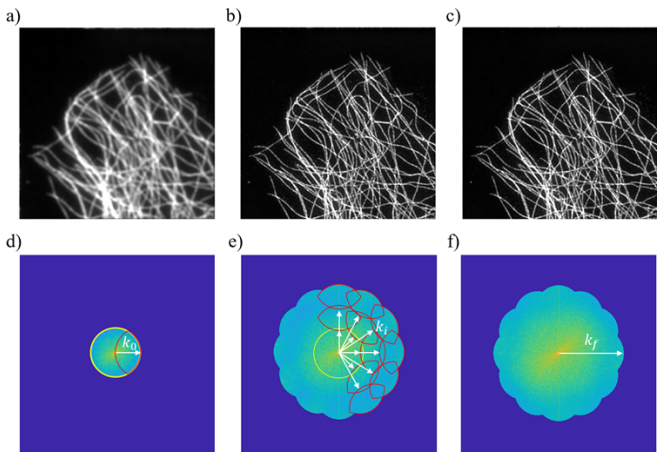

Fig. 4. **Cascading reconstruction simulation results**. Simulated diffraction limited U2OS grayscale image (511x511 pixels, uint8), cascading reconstruction results and the equivalent original high-resolution SIM image shown in real domain (a-c) and Fourier domain, in amplitude log scale, in (d-f) respectively. The yellow circle in (d-e) marks the 2D OTF region and the red ellipse-like shape marks the Fourier component region used for reconstruction. The white arrows in (d-f) depicts the cutoff frequency $k_0$, the illumination profiles frequencies $k_i$ and the maximal reconstructed frequency $k_f$. Illumination profile frequencies were chosen to provide continuous reconstruction of the Fourier domain using ellipse-like shapes.

The numerical simulation results shown in Fig. 4 provide strong confirmation to the proposed cascading reconstruction approach. Compared to the classical SIM mechanism, high-resolution images were accurately reconstructed, showcasing a practical cascading reconstruction process, achieved with a ~3 fold reduced number of frames and without modulating the phase of the illumination profile.

**Acknowledgments.** The authors acknowledge the fruitful discussions with S. Tsesses and O. Eyal.

**Disclosures.** The authors declare no competing financial interests.

**Data availability.** The data underlying the results presented in this paper are not publicly available but may be obtained from the authors upon reasonable request.

## REFERENCES

1. M. J. Rust, M. Bates, and X. Zhuang, "Stochastic optical reconstruction miscroscopy (STORM) provides sub-diffraction-limit image resolution," Nat. Methods **3**, 793–795 (2006).
2. E. Betzig, G. H. Patterson, R. Sougrat, O. W. Lindwasser, S. Olenych, J. S. Bonifacino, M. W. Davidson, J. Lippincott-Schwartz, and H. F. Hess, "Imaging Intracellular Fluorescent Proteins at Nanometer Resolution," Science (80-. ). **313**, 1642–1645 (2006).
3. S. W. Hell and J. Wichmann, "Breaking the diffraction resolution limit by stimulated emission: stimulated-emission-depletion fluorescence microscopy," Opt. Lett. **19**, 780 (1994).
4. M. G. L. Gustafsson, "Surpassing the lateral resolution limit by a factor of two using structured illumination microscopy.," J. Microsc. **198**, 82–7 (2000).
5. B. Yang, F. Przybilla, M. Mestre, J. Trebbia, and B. Lounis, "Large parallelization of STED nanoscopy using optical lattices," **22**, 11440–11445 (2014).
6. G. Moneron, R. Medda, B. Hein, A. Giske, V. Westphal, and S. W. Hell, "Fast STED microscopy with continuous wave fiber lasers," Opt. Express **18**, 1302 (2010).
7. W. Wang, H. Shen, B. Shuang, B. S. Hoener, L. J. Tauzin, N. A. Moringo, K. F. Kelly, and C. F. Landes, "Super Temporal-Resolved Microscopy (STReM)," J. Phys. Chem. Lett. **7**, 4524–4529 (2016).
8. P. W. Winter and H. Shroff, "Faster fluorescence microscopy: advances in high speed biological imaging," Curr. Opin. Chem. Biol. **20**, 46–53 (2014).
9. X. Chen, S. Zhong, Y. Hou, R. Cao, W. Wang, D. Li, Q. Dai, D. Kim, and P. Xi, "Superresolution structured illumination microscopy reconstruction algorithms: a review," Light Sci. Appl. **12**, 1–34 (2023).
10. Y. Blau, D. Shterman, G. Bartal, and B. Gjonaj, "Double moiré structured illumination microscopy with high-index materials," Opt. Lett. Vol. 41, Issue 15, pp. 3455-3458 **41**, 3455–3458 (2016).
11. D. Shterman, B. Gjonaj, and G. Bartal, "Experimental Demonstration of Multi Moiré Structured Illumination Microscopy," ACS Photonics **5**, 1898–1902 (2018).
12. F. Orieux, E. Sepulveda, V. Loriette, B. Dubertret, and J.-C. Olivo-Marin, "Bayesian estimation for optimized structured illumination microscopy.," IEEE Trans. Image Process. **21**, 601–14 (2012).
13. F. Ströhl and C. F. Kaminski, "Speed limits of structured illumination microscopy," Opt. Lett. **42**, 2511–2514 (2017).
14. A. Lal, C. Shan, K. Zhao, W. Liu, X. Huang, W. Zong, L. Chen, and P. Xi, "A Frequency Domain SIM Reconstruction Algorithm Using Reduced Number of Images," IEEE Trans. Image Process. **27**, 4555–4570 (2018).
15. M. G. L. Gustafsson, "Nonlinear structured-illumination microscopy: Wide-field fluorescence imaging with theoretically unlimited resolution," Proc. Natl. Acad. Sci. **102**, 13081–13086 (2005).